# Drive high power UVC-LED wafer into low-cost 4-inch era: effect of strain modulation

*Shangfeng Liu, Ye Yuan\*, Lijie Huang, Jin Zhang, Tao Wang, Tai Li, Junjie Kang, Wei Luo, Zhaoying Chen, Xiaoxiao Sun and Xinqiang Wang\**

S. Liu, T. Li, Dr. Z. Chen, Dr. X. Sun, Prof. X. Wang

State Key Laboratory for Mesoscopic Physics and Frontiers Science Center for Nano-optoelectronics, School of Physics, Peking University, Beijing 100871, China.

E-mail: wangshi@pku.edu.cn;

S. Liu, Dr. Y. Yuan, L. Huang, J. Zhang, Dr. J. Kang, Dr. W. Luo, Prof. X. Wang

Songshan Lake Materials Laboratory, Dongguan, Guangdong, 523808, China

E-mail: yuanye@sslab.org.cn;

Dr. T. Wang

Electron Microscopy Laboratory, School of Physics, Peking University, Beijing 100871, China

**Abstract**: Ultraviolet-C light-emitting diodes (UVC-LEDs) have great application in pathogen inactivation under various kinds of situations, especially in the fight against the COVID-19. Unfortunately, its epitaxial wafers are so far limited to 2-inch size, which greatly increases the cost of massive production. In this work, we report the 4-inch crack-free high-power UVC-LED wafer. This achievement relies on a proposed strain-tailored strategy, where a three-dimensional to two-dimensional (3D-2D) transition layer is introduced during the homo-epitaxy of AlN on high temperature annealed (HTA)-AlN template, which successfully drives the original compressive strain into tensile one and thus solves the challenge of realizing high quality $Al_{0.6}Ga_{0.4}N$ layer with a flat surface. This smooth $Al_{0.6}Ga_{0.4}N$ layer is nearly pseudomorphically grown on the strain-tailored HTA-AlN template, leading to 4-inch UVC-LED wafers with outstanding performances. Our strategy succeeds in compromising the bottlenecked contradictory in producing large-sized UVC-LED wafer on pronounced crystalline AlN template: The compressive strain in HTA-AlN allows for crack-free 4-inch wafer, but at the same time leads to a deterioration of the AlGaN morphology and crystal quality. The launch of 4-inch wafers makes the chip fabrication process of UVC-LEDs matches the mature blue one, and will definitely speed up the universal of UVC-LED in daily life.

## 1. Introduction

The explosion of COVID-19 has been greatly impacting the world and intensively activated the development of light-emitting diodes (LEDs) at the ultraviolet-C (wavelength ≤ 280 nm) emission range. It has been confirmed encouragingly perspective for the ultra-fast sterilization towards SARS-CoV-2 within one second[1-7]. In the past decades, in order to fabricate high-performance UVC-LED, various techniques have been proposed to seek excellent crystalline AlN templates on UVC-transparent sapphire substrates[8-19]. In spite that the epitaxial lateral overgrowth (ELOG)[4, 20-22] and high temperature annealing (HTA)[23, 24] strategies act as landmarks to fulfill the demands, both of them expose respective fatal shortcomings. Although 2-inch single-crystalline AlN templates with threading dislocation (TD) density down to ~$10^8$ cm$^{-2}$ is achieved by ELOG on nanopatterned sapphire substrate (NPSS), the lateral coalescing process produces intensive tensile strain which depressingly causes terrible cracks in 2-inch wafer[25]. On the other hand, from the viewpoint of industrial production, the complex preparation procedure of NPSS and necessary 3~4-μm-thick AlN layer for coalescing and dislocation annihilation[26, 27] unambiguously raise the cost. Recently, HTA is another highly admired technique for producing excellent crystalline AlN template due to its capability of reducing TD density down to $5\times10^8$ cm$^{-2}$ at a thickness less than 1 μm[23, 24]. And UVC-LEDs with wavelengths of 268 nm and 265 nm have been successfully fabricated on HTA-AlN[28, 29]. Moreover, the existence of compressive strain in HTA-AlN templates successfully suppresses cracks that happened on AlN/NPSS templates[25], thus illuminating the avenue towards 4-inch crack-free UVC-LED wafers, which directly matches the current mature industry process of GaN-based blue LED. Nevertheless, the exhibiting compressive-strain leads to serious surface roughening, lattice relaxation, and fresh-born TDs in the AlGaN epilayer especially when the Al mole fraction is 60% or lower[30-33]: The intensive compressive strain increases the surface diffusion barrier energy of Ga and Al adatoms, leading to hexagonal spiral island-growth of AlGaN layers along the screw- and mix-type dislocations which exhibit a component of the displacement vector normal to the growth surface[34-36]. Furthermore, due to the larger diffusion capability and incorporation efficiency of Ga adatoms than those of Al, compositional inhomogeneity occurs along the slope of the hexagonal hillocks, resulting in a degradation of device performance[35, 37]. Thus, the compressive strain caused morphology roughening yields a series of terrible problems in following UVC-LED epitaxy, acting as the main obstacle of HTA-AlN in UVC-LED fabrication.

Therefore, exploring a strategy solving crystalline quality, cracks, and surface roughening by controlling strain is of significant importance, especially for driving the high-performance UVC-LED into 4-inch size era which has never been approached in the field.

In this work, we initialize the high-performance 4-inch crack-free UVC-LED wafer. Through a strain-tailored strategy, i.e. introducing a three-dimensional to two-dimensional (3D-2D) transition layer on a 4-inch high-crystal-quality HTA-AlN template, the original compressive strain is successfully tuned into tensile one without sacrificing crystalline quality in the epitaxy process. As a result, crack and surface roughness are simultaneously suppressed. This work will promote the universalization of UVC-LED by utilizing low-cost 4-inch HTA-AlN templates particularly in terms of its compatibility with the current GaN-based blue LED process.

## 2. Results and discussion
### 2.1. Epitaxy and characterization of 4-inch AlN templates

The 4-inch HTA-AlN template was prepared via an ex-situ high temperature recrystallization process of 500-nm-thick AlN film deposited by physical vapor deposition (PVD). For comparison, a 4-inch AlN template on NPSS (AlN/NPSS) with hole-type pattern was grown by metal-organic chemical vapor deposition (MOCVD) through performing the ELOG technique presented in our previous work[2, 38]. Herein, the X-ray diffraction (XRD) rocking curves of (002) and (102) planes of HTA-AlN are shown in Fig. S4(a), with the full width at half maximums (FWHM) of 52 arcsec and 197 arcsec, respectively. According to the mosaic model, the total threading dislocation density is estimated as $5\times10^8$ cm$^{-2}$ [39], which is comparable to the crystalline quality of AlN/NPSS and is good enough to act as a promising platform to construct UVC-LED[33, 40]. Before UVC-LED epitaxy, 200-nm-thick AlN regrowth layer was homo-epitaxially grown by MOCVD on two kinds of AlN templates to ensure the fresh surface for subsequent UVC-LED epitaxy. In an attempt to develop an intuitive observation of our concerned surface-cracking situation, wafer-scaled mappings of the surface cracks were performed for the two types of AlN templates, where the green area denotes the crack region [Figs. 1(a) and 1(b)]. It is clearly shown that the AlN/NPSS endures terrible cracks in the wafer, as presented in Fig. 1(a). The quantitative count of cracks is 1607 pieces, as shown in Fig. S2. On these two AlN templates, a conventional UVC-LED structure was epitaxially grown, which consists of AlGaN buffer layer, *n*-AlGaN layer, multiple quantum wells (MQWs), electron blocking layer, and the *p*-AlGaN/*p*-GaN layers, as schematically shown in Fig. 5(a). Figure 1(c) presents the optical microscopy image of as-grown UVC-LED epilayer on AlN/NPSS template. The cracks in AlN/NPSS are reprinted in LED wafer, even in the central part of the wafer as shown in Fig. S3. Such a terrible crack is caused by the intensive tensile strain in the AlN template which is induced by the continuous AlN grain nucleation and merging during the ELOG process[27, 41], as schematically shown in Fig. 2(a). Compared with AlN on NPSS, AlN grown on flat sapphire substrate usually presents more cracks at the same thickness, as depicted in Fig. 2(b), because there is no naturally existing air void structure in the AlN epilayer to release tensile stress[33, 42]. Whereas for our 4-inch HTA-AlN template, both the high crystalline quality and the crack suppression are simultaneously compromised, as shown in Fig. 2(c): After the recrystallization process at high temperature, the quality of as-sputtered AlN film is improved by rearrangement of the AlN crystal lattice[43, 44]. Then in the cooling process, the thermo-mismatch between AlN and sapphire results in a compressive strain in AlN. As shown in Figs. 1(b) and 1(d), in contrast to the crack morphology on both AlN/NPSS and corresponding LED wafer, only a slight peeling happens at the edge region which is less than 0.5 mm away from the wafer boundary in both HTA-AlN template and LED epilayer. To quantitatively verify the different roles of the strain in the epitaxial layer, X-ray diffraction (XRD) reciprocal spacing mappings (RSM) of the AlN (105) plane were performed, and results are shown in Figs. 2(c) and 2(d), respectively. The strain-free AlN exhibits $Q_x$ and $Q_z$ values of 0.2858 and 0.7730, respectively, and the position is marked as white stars in the figures. According to the Bragg rule, $Q_x$ is in reverse proportion of the in-plane lattice constant $a$ [$Q_x = \lambda / (\sqrt{3}a)$, where $\lambda$=0.15406 nm is the wavelength of X-ray]. It is observed that the mapping peak position of HTA-AlN has larger $Q_x$ and smaller $Q_z$ values than those of strain-free AlN, indicating the existence of compressive strain. On the contrary, the AlN/NPSS presents smaller $Q_x$ and larger $Q_z$ values in comparison with those of strain-free AlN, suggesting the presence of tensile strain.

**2.2. UVC-LED fabrication on 4-inch HTA-AlN**

As mentioned, homo-epitaxial AlN layer was grown on HTA-AlN template by MOCVD and the morphology is recorded by atomic force microscopy (AFM). As shown in Fig. 3(b), a nice step bunching microscopic morphology with a root-mean-square roughness of 3 nm is observed. However, although the crystalline quality and surface morphology of HTA-AlN both satisfy the requirement for following AlGaN layer epitaxy, the compressive strain conceivably poses a tough challenge in the subsequent AlGaN growth: Due to the initial spiral steps provided by screw- and mixed-type dislocations and the larger diffusion mobility and incorporation efficiency of Ga adatoms at steps compared with Al adatoms, AlGaN layer presents hexagonal-hillock morphology with composition inhomogeneity[37, 45-47]; Such a phenomenon is further enhanced by compressive strain, leading to serious hexagonal island growth mode of *n*-type AlGaN layer and thus terrible surface roughening in following UVC-LED epitaxy[30, 33, 48-50], as shown in Fig. 3(a). It is worth noting that the growth mode of *n*-type AlGaN layer is the dominant factor to influence the quality of subsequent epitaxy. On one hand, the island-growth mode introduces tremendous dislocations into the *n*-AlGaN layer whose crystalline quality is continuously deteriorated upon increasing thickness. The freshly generated dislocations are harmful to the radiative recombination in multiple quantum wells (MQWs)[35, 51]. On the other hand, serious lattice relaxation takes place with surface roughening due to the compressive strain relaxation. Such a relaxation further decreases the transverse electric (TE)-polarized (which is perpendicular to *c*-plane) emission from the MQWs by modulating the valence bands of AlGaN[52]. Therefore, the light extraction efficiency is reduced. In our experiment, by referring to the 0% relaxation line (represented by white dashed line), the relaxation ratio is ~30% in *n*-AlGaN, as estimated from XRD RSM of (105) plane shown in Fig. 3(e). It is observed that the average intensity as well as the amplitude of the in-situ recorded reflectance at 405 nm both decrease rapidly as shown in Fig. 3(g), indicating surface deterioration. Embarrassingly, it seems that the advantage of compressive strain towards crack-suppression is becoming an obstacle when taking into account *n*-AlGaN morphology.

From the above discussion, we can conclude that strain control is the key point to solve the trade-off between crack generation and morphology degeneration of *n*-AlGaN. We then propose a 3D-2D transition layer on the HTA-AlN template to smoothen the surface of *n*-AlGaN epilayer as well as upper UVC-LED structure, i.e. a strain-tailored strategy. The 3D-2D transition layer growth process is schematically shown in Fig. 4(a) and the corresponding in-situ reflectance curve is shown in Fig. 4(b). During the 3D growth stage, a lot of AlN crystal grains homo-epitaxially grow on the HTA-AlN. Because the crystallographic orientation of these introduced AlN crystal islands is highly identical to that of HTA-AlN template, the outstanding crystalline quality of AlN is well kept. The 3D growth process results in surface roughening, therefore the surface reflectance decreases at this stage as shown in Fig. 4(b). During the subsequent 2D recovery process, the AlN crystal grains introduced in the 3D growth procedure tend to interconnect to reduce the effective area of surface, because the tensile strain energy created by the coalescing is smaller than the surface free energy of 3D island-like surface[41]. As a result, the recovery of surface flatness in the 2D growth stage successfully restores the surface reflectance intensity.

To confirm the crystal merging induced strain-modification, the (105) RSM of the strain-tailored HTA-AlN is performed as shown in Fig. 4(c). An obvious position shift of the AlN diffraction peak representing the tensile strain is shown and the pattern broadening does not show deterioration, which confirms that our growth strategy succeeds in reversing the strain state without obviously deteriorating the crystal quality. After adding the 3D-2D transition layer, different from the step bunching morphology of native HTA-AlN after regrowth, the surface of strain tailored HTA-AlN dramatically transforms to a step-flow morphology (root-mean-square roughness = 0.2 nm), as shown in Fig. 3(d). The crystalline quality of the strain tailored HTA-AlN is characterized by XRD and the FWHMs are 58 /237 arcsec for (002)/(102) plane rocking curves, respectively, as shown in Fig. S4(b). Such a tiny increase of FWHMs is negligible for the subsequent LED epitaxy. It is worth noting that, in conventional 3D AlN growth on sapphire, the lattice mismatch between AlN and sapphire causes lots of dislocations in the AlN islands and the inhomogeneous orientation of AlN grains induces new dislocations when the crystal grains merge. For our strain-tailored case, the homo-epitaxially grown 3D AlN grains have very consistent $c$-axis orientation and rotation angle [indicated by the parallel red arrows in Fig. 4(a)]. Hence tensile strain is induced without obviously deteriorating the crystal quality at the 2D coalescing stage.

Thanks to the strain modulation, a UVC-LED wafer with smooth surface is achieved, as shown in Fig. 3(c). The (105) plane RSM shown in Fig. 3(f) was performed to demonstrate the epitaxy quality of the $n$-AlGaN layer on strain tailored HTA-AlN[53, 54]. It is observed that the broadening of the $n$-AlGaN diffraction pattern with strain-tailor is close to that of the AlN pattern and is narrower than that of the sample without strain-tailor [Fig. 3(e)]. Accordingly, a relaxation degree of only 9% is estimated from the *n*-AlGaN diffraction pattern. Therefore, the *n*-AlGaN layer is nearly pseudomorphically grown and has excellent crystal quality. Figures 3(g) and 3(h) present the in-situ reflectance curves of the whole UVC-LED epitaxy process on the strain-tailored and native HTA-AlN templates, respectively. In Zone III shown in Fig. 3(h) (the *n*-AlGaN growth part), the reflectance curve of the strain-tailored sample shows a stable oscillation with a large average intensity, indicating 2D-growth and a flat surface of the *n*-AlGaN layer. Whereas the reflectance intensity for the native HTA-AlN continuously decreases to a small value, indicating surface roughening of the *n*-AlGaN layer [Fig. 3(g)]. As a result, with the aid of strain-tailored transition layer, we achieved an excellent UVC-LED structure with flat surface.

To demonstrate the following epitaxy quality, the high-angle annular dark-field scanning transmission electron microscopy (HAADF-STEM) was performed and the result is shown in Fig. 5(b) which focuses on the MQWs region. The shallow and dark regions represent quantum wells and barriers, respectively. It is clearly shown that the thicknesses of the quantum well and barrier are 2 and 11 nm, respectively. Within the detection limit, no visible dislocation is observed in the scan region. Moreover, the atomically sharp interface between the well and barrier regions indicates the excellent crystalline quality of MQWs[55].

Despite the strain-tailored HTA-AlN based 4-inch UVC-LED displays promising prospects, it exists shortage. Due to the strong compressive strain and large lattice and thermo-coefficient mismatch between epitaxial AlGaN/AlN and sapphire substrate, there exists a serious bow with a value as large as ~200 μm. The bowing results in an inhomogeneous temperature and flow field over the wafer during epitaxy in MOCVD, acting as a challenge to obtain homogeneous layer especially when wafer size is scaled up[33, 56]. Besides, the large bow leads to difficulties in the UVC-LED chip fabrication such as vacuum chuck handling and a curved focus plane in lithography[33]. Therefore, how to reduce the above-mentioned bow is regarded as one of the bottlenecks to match UVC LED fabrication with the conventional blue-LED chip process. Here, we propose a scratching idea on the backside of sapphire substrate and successfully reduce the bow down to ~150 µm by scratching a line as shown in Figs. S5 and S6. It is known that the umbrella shape bow of UVC-LED originates from thermal stress between the AlN/AlGaN epitaxial layer and the sapphire substrate during the cooling process of epitaxy. By introducing a scratch along the radial direction of the sapphire substrate, the created gap-space partially releases thermal stress. In spite that our scratch line is not completely cut through, the bow is partially reduced and the excellent AlN crystal quality (Fig. S7) is well kept. This strategy does do a great favor in improving the process and further promoting the HTA-AlN template in large UVC-LED wafers.

To demonstrate the uniformity of our UVC-LED wafer quantitatively, the sheet resistance and photoluminescence (PL) mapping were measured. As shown in Fig. 5(c), the results verify superior electrical uniformity of the UVC-LED wafer with an average sheet resistance of 95 $\Omega$/sq and an excellent wafer non-uniformity of ~1%. This promises the working voltage stability and electrical uniformity of LED chips within the wafer. Moreover, as described by the wafer-scaled wavelength mapping measured by PL shown in Fig. 5(d), excellent emission homogeneity is demonstrated by the small standard deviation of emission wavelength with a value of 2 nm at center emission wavelength of 275 nm. To intuitively test the electric-driven emission of the 4-inch UVC-LED wafer, the wafer-scale electroluminescence (EL) was measured and illustrated in Fig. 5(e): Four points were randomly selected to roughly examine the emission uniformity. It is seen that all points present UV-irradiation which qualitatively proves the homogeneity. Finally, figures 5(f) and 5(g) show the EL spectrum and current dependent output power of the fabricated UVC-LED chips on the HTA-AlN template. The chip size is 10 mil × 20 mil. At a forward current of 100 mA, an output power of 14 mW is obtained at the wavelength of 275 nm, where the FWHM is as small as 11 nm.

**3. Conclusion**

In conclusion, we initiated the crack-free 4-inch high power UVC-LED on HTA-AlN template through setting up a 3D-2D strain-tailor transition AlN layer. This transition layer solves the most challenging issue of HTA-AlN based LED epilayer: surface roughening induced by compressive strain. Moreover, the intensive wafer bowing induced by thermo-mismatch between the AlGaN and sapphire substrate is reduced by 25% through the strategy of "backside scratch". Our 4-inch UVC-LED will promote UVC-LED popularization by both reducing cost and enhancing productivity.

## 4. Experimental section

*Preparation of AlN templates:* The preparation of 4-inch high-quality AlN template was carried out by combing sputtering with high temperature face-to-face annealing. C-plane sapphire with miscut angles of c to m 0.2±0.1° and c to a 0±0.1° was used as the substrate. During the sputtering process, pure aluminium (99.999%) was used as the sputtering target. The sputtering power and temperature were set as 3000 W and 550 °C, respectively. The mixture of argon and nitrogen was the sputtering atmosphere with the volume ratio of 1:4. Calibrated by ellipsometry, the thickness of the obtained layer was determined as 500 nm. During the annealing operation, a specific face-to-face operation was used, and the annealing ambient was set as nitrogen (99.99%). The annealing condition was set as 1700 °C for five hours. Subsequently, the sample was cooling down to room temperature under nitrogen ambient. The AlN films on NPSS are epitaxially grown in Prismo HiT3$^{TM}$ MOCVD system. $H_2$ and $N_2$ were used as the carrier gas for the epitaxial process. Trimethyl-aluminum (TMAl) and ammonia ($NH_3$) were used as Al and N precursors, respectively. Prior to the epitaxy of AlN films, a 15-nm-thick AlN layer was deposited on NPSS by magnetron sputtering as a nucleation layer. The AlN epitaxy growth on NPSS is consisted of three stages: (i) a 200-nm-thick layer growth in three-dimensional (3D) mode (temperature = 1100 °C, pressure = 100 Torr); (ii) the lateral overgrowth with a thickness of 1.8 μm (temperature = 1250 °C, pressure = 40 Torr); (iii) a 2-μm continuing growth (temperature = 1200 °C pressure = 30 Torr).

*Epitaxial growth of UVC LED on HTA AlN template:* UVC LED structure epitaxy was grown on the obtained HTA AlN templates by Prismo HiT3$^{TM}$ MOCVD system. Trimethylgallium (TMGa), trimethylaluminum (TMAl), and ammonia ($NH_3$) were used as Ga, Al, and N precursors, respectively. $H_2$ and $N_2$ were used as carrier gases. $SiH_4$ and CpMg were used as the doping source of *n*-type and *p*-type AlGaN. For the 3D-2D transition layer, the AlN 3D growth was performed at a relative lower temperature = 1100 °C, chamber pressure = 100 Torr, V/III = 2010 and the 2D AlN coalescing process was performed at growth temperature = 1250 °C, chamber pressure = 30 Torr, V/III = 600; The growth continued with a 140 nm thick $Al_{0.8}Ga_{0.2}N$ buffer layer and 1100 nm thick Si-doped *n*-$Al_{0.60}Ga_{0.40}N$ layer at 1100 °C. The subsequent MQWs region consists of 5 pairs of 2 nm thick $Al_{0.43}Ga_{0.57}N$ quantum well layers and 11 nm thick $Al_{0.52}Ga_{0.48}N$ quantum barrier layers. Subsequently, a Mg-doped 15 nm thick $Al_{0.8}Ga_{0.2}N$ layer was grown on the MQW region as an electron blocking layer (EBL). The growth was terminated with a *p*-$Al_{0.6}Ga_{0.4}N$ layer and a subsequent 20 nm *p*-GaN layer. After epitaxial growth, the standard UVC LED chip processing was performed by mask layer deposition, photolithography, reactive ion etching, and sputtering techniques to fabricate 10 mil × 20 mil chips.

*Characterization:* The surface crack mappings of the 4-inch wafer were measured by KLA-Tencor Candela CS20. The crystalline quality and RSM patterns of AlN templates and UVC LED structures were characterized by PANalytical X'Pert3 MRD XL system operating at 40 mA and 45 kV using Cu Kα1 radiation (0.154056 nm). The sheet resistance mapping was measured by Semilab LEI-1510 to demonstrate the uniformity of electrical properties. The PL mapping of the UVC LED wafer was performed by Etamax PLATO PL system including a laser (216 nm) as an excitation source.

Acknowledgments

This work was partly supported by Beijing Outstanding Young Scientist Program (No. BJJWZYJH0120191000103), Guangdong Basic and Applied Basic Research Foundation (No. 2020A1515110891), the National Natural Science Foundation of China (Nos. 61734001). Many appreciations for the fruitful discussion with Dr. Long Yan, Dr. Jason Hoo, and Dr. Shiping Guo from Advanced Micro-Fabrication Equipment Inc (AMEC).

Received: ((will be filled in by the editorial staff))
Revised: ((will be filled in by the editorial staff))
Published online: ((will be filled in by the editorial staff))

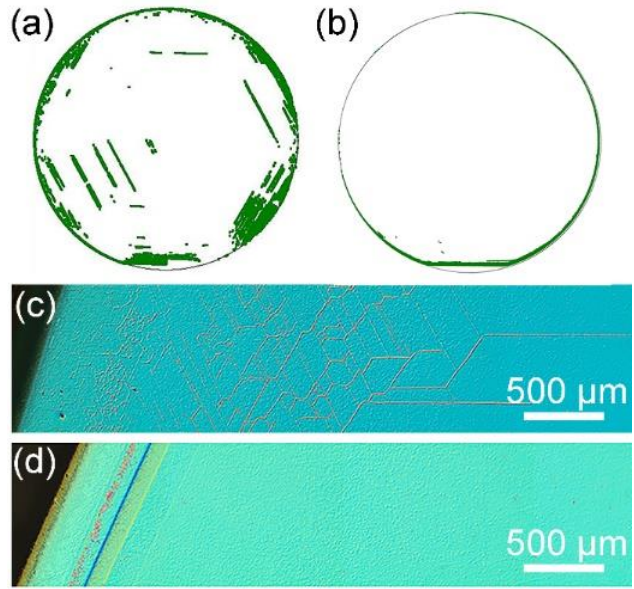

**Figure 1.** (Color online) The crack distributions in (a) 4-inch AlN/NPSS and (b) 4-inch HTA-AlN templates, as well as the corresponding optical microscopy images of LED wafers based on (c) 4-inch AlN/NPSS and (d) 4-inch HTA-AlN templates.

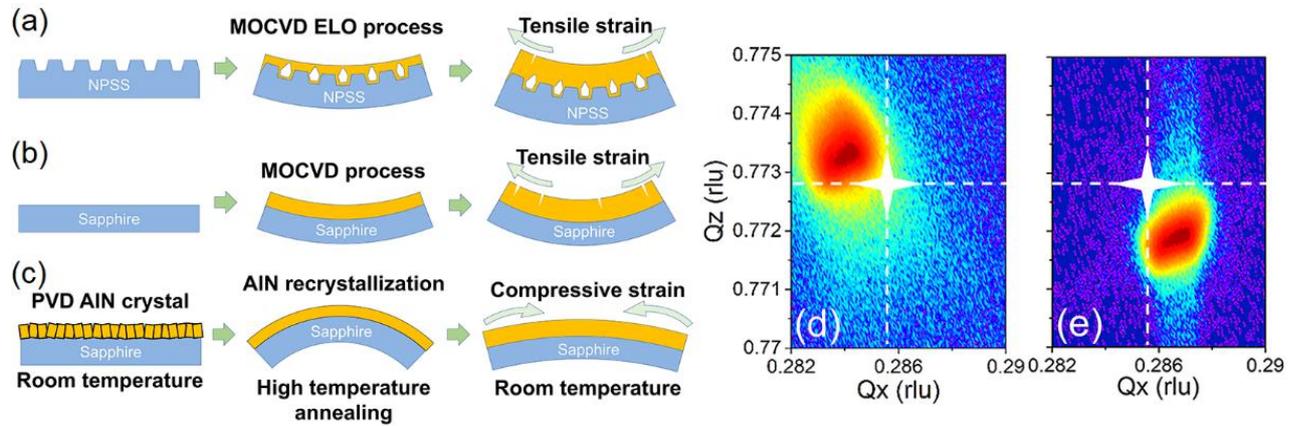

**Figure 2**. (Color online) The scheme of (a) AlN/NPSS; (b) AlN/Flat Sapphire substrate; (c) HTA-AlN template preparation which present different types of inner-strain; The X-ray diffraction of (105) plane RSMs of (d) AlN/NPSS and (e) HTA-AlN. The diffraction peak of strain-free (105) plane AlN is star-marked.

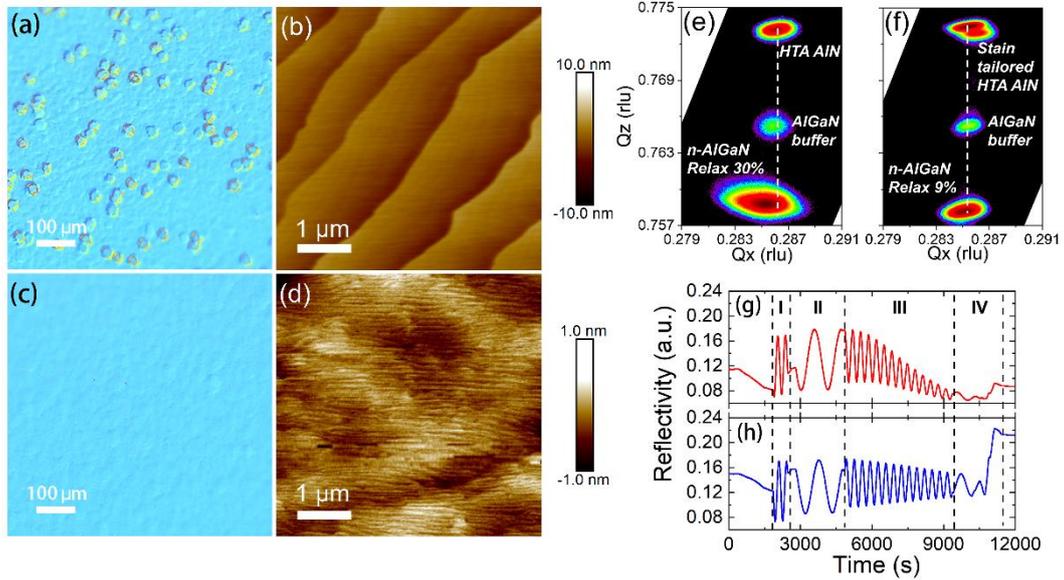

**Figure 3.** (Color online) The optical microscopy images of the UVC LED wafers on HTA AlN (a) without and (c) with 3D-2D transition layer. The atomic force microscopy images of the HTA-AlN after AlN regrowth (b) without and (d) with 3D-2D transition layer. The X-ray (105) plane RSMs of UVC-LED structure on HTA-AlN templates (e) without and (f) with 3D-2D transition layer. The diffraction pattern broadening is generally considered a distinctive reference for the quality of AlGaN layer. From the peak positions of *n*-AlGaN and AlN RSM, the relaxation ratios are calculated to be 30% and 9%. The in-situ 405 nm reflectance curves of the UVC-LED grown on HTA-AlN (g) without and (h) with 3D-2D transition layer. The corresponding different stages in the UVC-LED epitaxy are marked with dashed lines: Zone I (AlN regrowth), Zone II (AlGaN transition layer), Zone III (*n*-AlGaN layer), and Zone IV (MQW and *p*-type region).

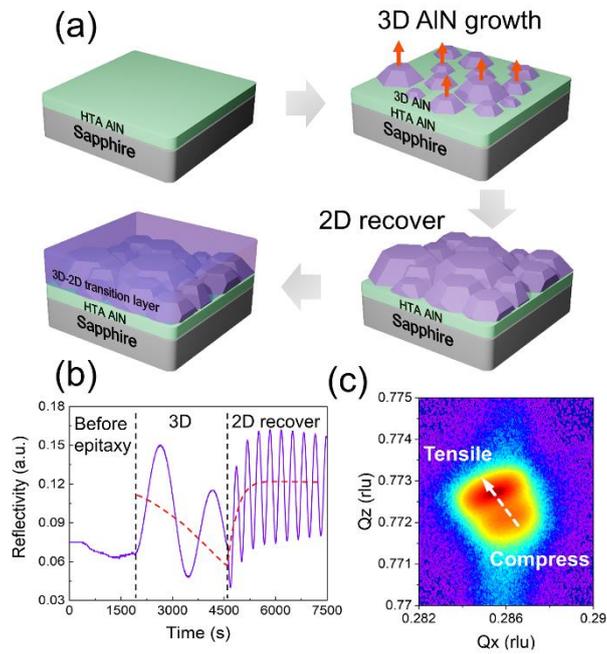

**Figure 4.** (Color online) (a) Schematic and (b) corresponding in-situ 405 nm reflectance curve of the 3D-2D transition layer growth process on HTA-AlN. The red dashed line is a guideline for eyes to catch the trend of the average intensity of the reflectance curve. (c) The X-ray (105) plane RSM of strain tailored HTA-AlN template. The trace of strain shift is marked by the white dashed arrow.

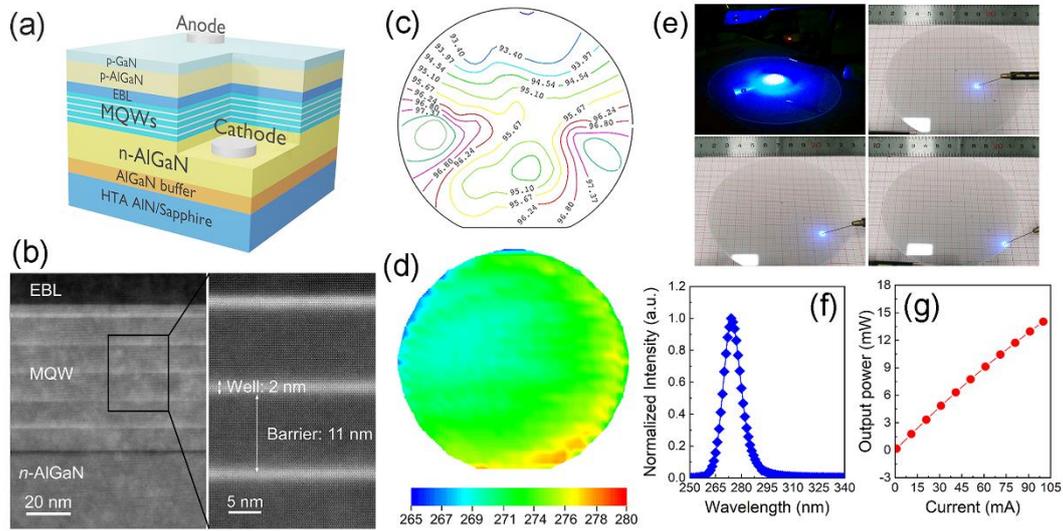

**Figure 5.** (Color online) (a) Schematic diagram of the UVC-LED structure prepared by MOCVD on 4-inch strain-tailored HTA-AlN template; (b) HAADF-STEM images of the UVC-LED as well as MQWs region. It is observed that the 5-period MQW region consists of 2-nm-thick $Al_{0.43}Ga_{0.57}N$ wells and 11-nm-thick $Al_{0.52}Ga_{0.48}N$ barriers; (c) The sheet resistance mapping (unit: Ω/sq) and (d) PL wavelength mapping (unit: nm) of the 4-inch UVC LED wafer on strain-tailored HTA-AlN template; (e) The photographs of EL of the as-grown UVC-LED wafer on strain-tailored HTA-AlN template; (f) The wavelength-dependent EL and (g) output power as a function of current for flip-chip UVC-LED on strain-tailored HTA-AlN template.

**Table of Contents**

We propose a strain-tailoring strategy in AlN epitaxy on high temperature annealed AlN template and thus demonstrate the first 4-inch crack-free high power UVC-LED wafer. Such a success pushes UVC-LEDs into the same 4-inch area as commercially mature blue LEDs, enabling the possibility of large-scale epitaxy and chip manufacturing at significantly reduced cost.

**Keywords:** AlN, 4-inch, high temperature annealing, Ultraviolet-C LED, strain modulation

**Authors:** Shangfeng Liu, Ye Yuan*, Lijie Huang, Jin Zhang, Tao Wang, Tai Li, Wei Luo, Junjie Kang, Zhaoying Chen, Xiaoxiao Sun and Xinqiang Wang*

**Title:** Drive high power UVC-LED wafer into low-cost 4-inch era: effect of strain modulation

**ToC Figure**

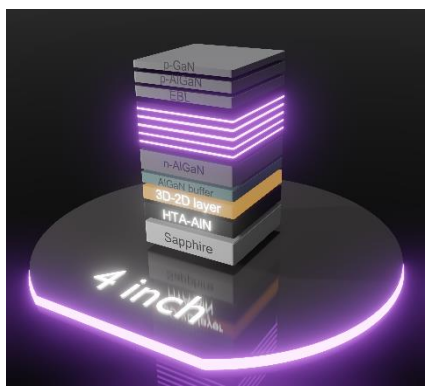